\documentclass[letterpaper,conference]{IEEEtran}
\IEEEoverridecommandlockouts
\usepackage{graphicx}
\usepackage[utf8]{inputenc} 
\usepackage[T1]{fontenc}
\usepackage{url}
\usepackage{ifthen}
\usepackage{cite}
\usepackage{color}

\usepackage{cite}

\ifCLASSINFOpdf
\else
\fi
\usepackage{array}
\usepackage{graphicx}
\usepackage{hyperref}
\usepackage{amsmath}
\usepackage{amssymb}
\usepackage{bm}
\usepackage{bbm}
\usepackage{algorithmicx}
\usepackage{algorithm}
\usepackage{algpseudocode}
\usepackage{array}
\usepackage{booktabs}
\usepackage{multirow}
\usepackage{makecell}
\usepackage{tikz}
\usetikzlibrary{shapes.geometric, arrows}

\begin{document}
%

\title{CRC-Aided Short Convolutional Codes and RCU Bounds for Orthogonal Signaling }



%

\author{\IEEEauthorblockN{Jacob King\IEEEauthorrefmark{1}\IEEEauthorrefmark{2},
William Ryan\IEEEauthorrefmark{2}, and
Richard D. Wesel\IEEEauthorrefmark{1}
}
\IEEEauthorblockA{\IEEEauthorrefmark{1}Department of Electrical and Computer Engineering, University of California, Los Angeles, Los Angeles, CA 90095, USA}
\IEEEauthorblockA{\IEEEauthorrefmark{2}Zeta Associates, Aurora, Denver, CO 80011, USA}
Email: jacob.king@ucla.edu, ryan-william@zai.com, wesel@ucla.edu,}


\maketitle

\begin{abstract}
We extend earlier work on the design of convolutional code-specific CRC codes to $Q$-ary alphabets, with an eye toward $Q$-ary orthogonal signaling. Starting with distance- spectrum optimal, zero-terminated, $Q$-ary convolutional codes, we design $Q$-ary CRC codes so that the CRC/convolutional concatenation is distance-spectrum optimal. The $Q$-ary code symbols are mapped to a $Q$-ary orthogonal signal set and sent over an AWGN channel with noncoherent reception.  We focus on $Q=4$, rate-1/2 convolutional codes in our designs.  The random coding union bound and normal approximation are used in earlier works as benchmarks for performance for distance-spectrum-optimal convolutional codes.  We derive a saddlepoint approximation of the random coding union bound for the coded noncoherent signaling channel, as well as a normal approximation for this channel, and compare the performance of our codes to these limits. Our best design is within $0.6$ dB of the RCU bound at a frame error rate of $10^{-4}$.

\end{abstract}


%
\IEEEpeerreviewmaketitle
{\let\thefootnote\relax\footnote{{This research is supported by Zeta Associates Inc. and National Science Foundation (NSF) grant CCF-2008918. Any opinions, findings, and conclusions or recommendations expressed in this material are those of the author(s) and do not necessarily reflect views of Zeta Associates Inc. or NSF.}}}

\section{Introduction}
\label{sec:Intro}

\subsection{Background}

Phase coherency between transmitter and receiver is necessary for optimal reception.  However, phase coherency can be difficult to achieve in practice, so orthogonal signaling with noncoherent reception is often used.  The most common examples of orthogonal signal sets are $Q$-ary Hadamard sequences and $Q$-ary frequency shift keying (QFSK) \cite{Proakis}.  We will assume the latter throughout this paper.  Non-coherent FSK signaling is of practical importance.  It is currently used in Bluetooth \cite{Bluetooth}.  More recently the LoRa standard has adopted noncoherent QFSK signaling \cite{Lora} \cite{BLE2LoRa}.

For values of $Q$ greater than 8, noncoherent QFSK loss is small compared to coherent QFSK.  In addition, for large values of $Q$, noncoherent QFSK performs nearly as well as BPSK signaling, at the expense of bandwidth.  With these facts in mind, developing good codes for noncoherent QFSK is very important for contexts in which phase coherency is difficult or impossible. This occurs when there is a high relative velocity between the transmitter and the receiver or when the receiver must be very simple or inexpensive.
A natural code choice for QFSK is a code based on a $Q$-ary alphabet so that
code symbols are directly mapped to modulation symbols.

Binary convolutional codes concatenated with binary CRC codes have been shown to perform very well on BPSK/QPSK channels \cite{Lou2015} \cite{YangTBCC} \cite{Yang2022}. Following \cite{Lou2015}, we design $Q$-ary cyclic redundancy check (CRC) codes to be concatenated with optimal, $Q$-ary, zero-state-terminated convolutional codes (ZTCC), where zeros are appended to the end of the CRC word to force the convolutional encoder to terminate in the zero state.  We denote this concatenated code by CRC-ZTCC.  

The $Q$-ary CRC code design criterion is optimization of the distance spectrum of the concatenation of the CRC code represented by $g(x)$ and the convolutional code represented by $[g_1(x)\,\ g_2(x)]$, where each polynomial has $Q$-ary coefficients. With all operations over GF($Q$), this concatenation is equivalent to a  $Q$-ary convolutional code with polynomials $[g(x)g_1(x)\,\ g(x)g_2(x)]$, which is ostensibly a catastrophic convolutional code. However, rather than applying a Viterbi decoder to this resultant code, we employ the list Viterbi algorithm (LVA) \cite{Seshadri1994}.  The LVA produces a list of candidate trellis paths in the original convolutional code trellis, ordered by their likelihoods, and then chooses as its decision the most likely path to pass the CRC check.

Design of codes for noncoherent orthogonal signaling has been done for long messages in \cite{Fabregas, Valenti, Valenti2, Valenti3}.  Here, we analyze $Q$-ary CRC-ZTCC codes for short messages. Optimal $Q$-ary convolutional codes for orthogonal signaling were described by Ryan and Wilson \cite{Ryan}.  We design distance-spectrum optimal (DSO) CRCs for two of the codes in \cite{Ryan}. 

Since the pioneering work of\ Polyanskiy \emph{et al.} \cite{Polyanskiy}, the random coding union (RCU) bound has been used as a measure of the performance quality of short-message binary codes. The RCU bound is very difficult to calculate, but   Font-Segura \emph{et al.} \cite{Saddlepoint}  derived a saddlepoint approximation for the RCU bound that is more practical to calculate.  In this paper we extend their work to the noncoherent QFSK channel.  We also include here the normal approximation to the RCU bound for its simplicity. A converse sphere packing bound was also presented by Shannon \cite{Shannon} as a lower bound on error rate for finite blocklength codes, and revisited by \cite{Valembois}.

\subsection{Contributions}

This paper designs DSO $Q$-ary CRC codes for two 4-ary ZTCCs selected from \cite{Ryan} and we apply their concatenation to the noncoherent 4-FSK channel with list Viterbi decoding.  We also derive a saddlepoint approximation of the RCU bound for the special case of the noncoherent QFSK channel.  The performances of the codes designed are compared to their respective RCU bounds and the normal approximation.  Applying these techniques to larger values of $Q$ is an area for future work.

\subsection{Organization}

Section \ref{sec:Channel} details the channel model and properties of the noncoherent QFSK channel.  Section \ref{sec:CodeDesign} then describes the design criteria for optimal CRC-ZTCC concatenated codes, as well as the algorithm for finding optimal CRCs.  Section \ref{sec:RCU} then shows the equations for the saddlepoint approximation for the RCU bound and derives the relevant equations for the noncoherent QFSK channel.  Finally, Section \ref{sec:Results} presents the performance of optimal CRC-ZTCC codes compared to the RCU bound.

\section{Channel Model}
\label{sec:Channel}

Our discussion here of the noncoherent QFSK channel follows\cite{Wilson}.  For a message symbol ${x \in \{1, 2, ..., Q\}}$, the transmitter takes $x = i$ and transmits the corresponding duration-$T$ signal $s_i(t, \phi) = A \cos (\omega_i t + \phi)$, where ${A = \sqrt{2E_s/T}}$ so that the energy of the signal is $E_s$, $\phi$ is uniform over $[0, 2\pi)$, and the frequencies $\omega_i/2\pi$ are separated by a multiple of the symbol rate to ensure mutual orthogonality among the signals $s_i(t)$.  

The detector receives the signal $r(t) = s_i(t, \phi) + n(t)$, where $n(t)$ is zero mean AWGN with power spectral density $N_0/2$.
The detector consists of $Q$ pairs of correlators, with the $j$th pair correlating $r(t)$ against $\frac{2}{N_0} s_j(t, 0)$ and $\frac{2}{N_0} s_j(t, \frac{\pi}{2})$.  The two correlator outputs are then squared and summed, and a square root is taken of the result.  We denote this root-sum-square of the two values by $y_j$.  The vector $y = [y_1, ..., y_Q]^T$ is the soft decision output of the detector.

If $i \ne j$, the correlation of $s_i(t, \phi)$ and $s_j(t, 0)$ is 0 due to orthogonality, and the same is true for the correlation of $s_i(t, \phi)$ and $s_j(t, \frac{\pi}{2})$.  As such, the value $y_j$ will be the root-sum-square of two zero-mean Gaussian random variables with variance $\sigma^2 = 2E_s/N_0$.  Thus, $y_j$ will have a Rayleigh distribution with parameter $\sigma^2 = 2E_s/N_0$.  If $i = j$, however, the Gaussian random variables that are root-sum-squared will not be zero mean.  As a result, $y_j$ will instead have a Rice distribution with parameters $\mu = 2E_s/N_0$ and $\sigma^2 = 2E_s/N_0$.  The Rayleigh and Rice distributions are as follows:

\[
    f_{\mathrm{Rayleigh}}(y_j\,|\,x=i) = \frac{y_j}{\sigma^2} \exp\left[- \frac{y_j^2}{2\sigma^2}\right] \tag{1}
\]

\[
    f_{\mathrm{Rice}}(y_i\,|\,x=i) = \frac{y_i}{\sigma^2} I_0(y_i) \exp \left[- \frac{y_i^2 + \mu^2}{2 \sigma^2}\right] \tag{2}
\] where $I_0(.)$ is the zeroth-order modified Bessel function of the first kind.

The optimal decoding metrics, i.e., log likelihoods, for the AWGN channel involves the logarithm of a Bessel function \cite{Wilson} which is clearly impractical. In practice, the "square-law metric" $y_i^2$ is generally used instead of the optimal metrics \cite{Fabregas, Ryan, Stark}. We found that $y_i$ performs better than the square-law metric $y_i^2$ suggested in these papers, and $y_i$ is a better approximation for the optimal metric.  Optimal decoding for noncoherent QFSK is an area for future attention.

Given the message symbol $x = i$, the received vector $y$ has a density function that is the product of one Rice density function, corresponding to $y_i$, and $Q-1$ Rayleigh density functions, corresponding to all $y_j$ for $j \ne i$.  This yields the following transition probabilities for the noncoherent QFSK channel:

\[
    W(y \, | \, x = i) = \frac{\prod_{k = 1}^Q y_k}{\sigma^{2Q}} I_0(y_i) \exp \left[ {- \frac{\mu^2 + \sum_{k = 1}^Q y_k^2}{2\sigma^2}} \right ] \tag{3}
\]

\section{CRC/Convolutional Code Design for QFSK}
\label{sec:CodeDesign}

The asymptotic (in signal-to-noise ratio, SNR) codeword-error rate (also, frame-error rate) for a length-$n$ block code with minimum distance $d_{min}$ on the QFSK/AWGN channel is union upper bounded \cite{Ryan} as 

\[
    P_{cw} < \sum^{n}_{d=d_{min}}\ N(d)P_2(d) \tag{4}
\]
where $N(d)$ is the number of weight-$d$ codewords in the code
and $P_2(d)$ is the pairwise error probability for two codewords at
distance $d$. Asymptotically in SNR, $P_2(d)$ decreases with increasing
$d$ \cite{Ryan} so that, from the bound, codes should be designed with   $d_{min}$ as large as possible. Also from the bound, for each $d$, the
multiplicities $N(d)$ should be as small as possible. Codes satisfying these criteria are called \emph{distance-spectrum optimal} (DSO).

Ryan and Wilson \cite{Ryan} have found optimal non-binary convolutional codes for small memory.  These codes are optimal in the sense of maximizing the free distance $d_{free}$ and minimizing the information symbol weight at each weight $w \ge d_{free}$.

In 2015, Lou \emph{et. al.} \cite{Lou2015} showed the importance of designing CRC codes for specific convolutional codes.  An optimal CRC should minimize the frame error rate (FER) of the CRC-ZTCC concatenated code based on the union bound above on FER.  These CRCs are called distance-spectrum-optimal CRCs.  It can also be shown that, at high signal-to-noise ratios, this is equivalent to maximizing the minimum Hamming distance $d_{min}$ of the concatenated code, and minimizing the number of codewords $N(d_{min})$ at $d_{min}$ and weights near $d_{min}$.  This criterion is very similar to the criterion for the optimal convolutional codes in \cite{Ryan}.

In this paper, we adapt the methods in \cite{Lou2015} to find DSO CRCs for 4-ary convolutional codes.  We consider a memory-2 ($\nu = 2$) and a memory-4 ($\nu = 4$) code presented in \cite{Ryan}.  The convolutional code generator polynomials $g_1$ and $g_2$ can be found in Table \ref{tab:gen} and the optimal CRC polynomials are in Table \ref{tab:CRC}, with the $x^0$ coefficient appearing on the left.  These polynomials are elements of $GF(4)[x]$, with $GF(4) = \{0, 1, \alpha, \beta\}$ where $\alpha$ is a primitive element of $GF(4)$ and $\beta = \alpha^2$.

\renewcommand{\tabcolsep}{3pt}

\begin{table}[]
    \centering
    \caption{Generator Polynomials for the Memory-2 and Memory-4 4-ary Convolutional Codes}
    \begin{tabular}{|c|c|c|c|c|c|}
        \hline
        $\nu$ & $g_1$ & $g_2$ & $d_{free}$ & $N_t(d_{free})$ & $N_c(d_{free})$  \\
        \hline
        2 & (1, 1, 1) & (1, $\alpha$, 1) & 6 & 6 & 381 \\
        \hline
        4 & (1, 1, 1, $\beta$, $\alpha$) & (1, $\alpha$, 1, $\alpha$, $\beta$) & 9 & 6 & 378 \\
        \hline
    \end{tabular}
    
    \label{tab:gen}
\end{table}

\begin{table}[]
    \centering
    \caption{DSO CRC Polynomials for the Memory-2 and Memory-4 4-ary Convolutional Codes}
    \begin{tabular}{|c|c|c|c|c|c|}
        \hline
        $\nu$ & $m$ & $g$ & $d_{min}$ & $N_{t}(d_{min})$ & $N_c(d_{min})$  \\
        \hline
        2 & 3 & (1, $\beta$, 1, $\alpha$) & 11 & 21 & 1305 \\
        \hline
        2 & 4 & (1,  0, 0, $\beta$, $\alpha$) & 12 & 18 & 612 \\
        \hline
        2 & 5 & (1, 0, 0, $\alpha$, $\beta$, 1) & 13 & 6 & 273 \\
        \hline
        2 & 6 & (1, $\alpha$, $\beta$, 0, 1, 1, $\alpha$) & 15 & 48 & 2442 \\
        \hline
        2 & 7 & (1, 0, 1, $\beta$, $\beta$, $\beta$, 0, $\alpha$ ) & 16 & 21 & 1029 \\
        \hline
        2 & 8 & (1, $\alpha$, $\alpha$, $\alpha$, 1, $\alpha$, $\alpha$, $\alpha$, $\beta$) & 17 & 9 & 345 \\
        \hline
        4 & 3 & (1, $\beta$, $\alpha$, $\beta$) & 14 & 30 & 1839 \\
        \hline
        4 & 4 & (1, 0, 0, $\beta$, $\beta$) & 15 & 15 & 921 \\
        \hline
        4 & 5 & (1, 0, $\beta$, $\beta$, 0, 1) & 16 & 3 & 174 \\
        \hline
        4 & 6 & (1, $\beta$, 1, $\alpha$, $\alpha$, 1, $\beta$) & 18 & 21 & 1266 \\
        \hline
        4 & 7 & (1, 1, 1, $\alpha$, $\beta$, $\beta$, 1, $\alpha$) & 19 & 9 & 561 \\
        \hline
    \end{tabular}
    
    \label{tab:CRC}
\end{table}

The DSO CRC polynomials for each convolutional code and each CRC-ZTCC are found through an exhaustive search.  We begin by initializing a list with every CRC polynomial of degree $m$ and setting a max weight to search to $\tilde{d}$.  For every weight from $w = d_{free}$ to $w = \tilde{d}$, we find the number of codewords of Hamming weight $w$ for each CRC-ZTCC concatenated code with polynomials $[g(x)g_1(x) \,\ g(x)g_2(x)]$.  

Codewords are found by the same process used in \cite{Lou2015}, adapted for CRC-ZTCC codes in $GF(4)$.  This is done by traversing through the trellis of the CRC-ZTCC code for each CRC.  We begin in the zero state.  For 4-ary CRC-ZTCC codes, each state can transition into four possible new states, one for each element of $GF(4)$.  We traverse through the trellis, allowing all possible state transitions, and we maintain a list of every codeword constructed this way.  A trellis path is eliminated from contention if the corresponding codeword reaches a weight of $\tilde{d}$ before rejoining the zero state.  If a path reaches the end of the trellis in the zero state with a codeword weight $w \le \tilde{d}$, we increment the count of the number of codewords at weight $w$ for this CRC-ZTCC.

After the distance spectra for every CRC-ZTCC is found, we find which CRC-ZTCC has the largest $d_{min}$.  If multiple CRC-ZTCCs have the same $d_{min}$, we select whichever CRC-ZTCC has the least number of codewords at $d_{min}$.  If there is a tie for the smallest number of codewords at $d_{min}$, we compare the number of codewords at $d_{min} + 1$, and we continue incrementing until the tie is broken.  Table \ref{tab:gen} and Table \ref{tab:CRC} show the minimum distances $d_{free}$ and $d_{min}$ for the convolutional codes and CRC-ZTCCs, respectively.

Often, the distance spectrum for a convolutional code is given in terms of of the number of error events at each weight $w$, as in \cite{LinCostello}.  This metric only cares about the number of paths on the trellis that diverge from the zero state and eventually rejoin, independent of codeword length.  However, in this paper we analyze CRC-ZTCCs as a block code, so the more important metric is the number of codewords of weight $w$. Table \ref{tab:gen} and Table \ref{tab:CRC} provide both the number of error events on the trellis at $w = d_{min}$, $N_t(d_{min})$, and the number of codewords for the block code, $N_c(d_{min})$.

\begin{figure*}[t]
    \[
    g'(y, \rho) = g(y, \rho) \left (\log f(y, \rho) + \frac{(1+\rho)f'(y, \rho)}{f(y, \rho)} \right ) + f(y, \rho)^{1 + \rho} \left (\frac{2 f'(y, \rho)}{f(y, \rho)} + (1 + \rho) \left ( \frac{f''(y, \rho)}{f(y, \rho)} - \left (\frac{f'(y, \rho)}{f(y, \rho)} \right)^2 \right ) \right ) \tag{16} \label{eq:gprime}
    \]
    \centering
    \hrulefill
\end{figure*}

\section{RCU Bound Equations}
\label{sec:RCU}

The RCU bound is an achievability bound for codes of a given rate and finite blocklength, first described by Polyanskiy, Poor, and Verd\'u in 2010 \cite{Polyanskiy}. The RCU bound is defined in \cite{Polyanskiy} as follows: let $n$ and $m$ be positive integers.  Let $P^n(x)$ be a probability distribution for a random coding ensemble for codewords of length $n$, and let $W^n(y|x)$ be a length-$n$ channel transition probability.  The RCU bound for a length-$n$ code with $M$ codewords is given by

\[
\mathrm{rcu}(n, M) = \mathbb{E}_{X, Y} \left [\min \left \{1, (M-1) \text{pep}(X,Y)  \right\} \right ] \tag{5} \label{eq:polyanskiy}
\] where $\mathbb{E}_{X, Y}$ is the expectation over $X$ and $Y$, $X$ is a random variable drawn from $P^n(x)$, $Y$ is a random variable drawn from $W^n(y|X)$, 

\[
\mathrm{pep}(X,Y) = \mathbb{P}[i(\bar{X};Y) \ge i(X;Y) \; | \; X, Y] \tag{6} \label{eq:pep}
\] is the pairwise error probability with $\bar{X}$ drawn from $P^n(x)$, and $i(X;Y)$ is the mutual information density of $X$ and $Y$.

Calculating the RCU bound using this definition is computationally hard for most practical situations.  In 2018, Font-Segura \emph{et. al.} \cite{Saddlepoint} presented a saddlepoint approximation for the RCU bound to reduce computation complexity.  In this section, we will present the equations for the saddlepoint approximation of the RCU bound, find expressions for the derivatives of necessary functions, and apply the noncoherent orthogonal signal channel model to the equations in \cite{Saddlepoint}.

We start with Gallager's $E_0$-function \cite{Gallager}, which is a function of a distribution over the message symbol alphabet $P(x)$ and channel $W(y \, | \, x = i)$.  We will write $W(y \, | \, i)$ for $W(y \, | \, x = i)$ for notational simplicity.  The Gallager $E_0$ function is defined as

\[
E_0(\rho) = -\log \int \left (\sum_{i=1}^Q P(x=i) W(y \, | \, i) ^{\frac{1}{1+\rho}} \right )^{1+\rho} \mathrm{d}y \tag{7} \label{eq:E0}
\] where $\log$ is the natural logarithm.  For the saddlepoint approximation of the RCU bound, we must find the first and second derivatives of $E_0(\rho)$ with respect to $\rho$.  We will assume a uniform distribution $P(x) = 1/Q$, as this is optimal for symmetric channels as is the case for our channel.

These derivatives are notationally complex due to exponentiation in $\rho$, so to simplify we define the following functions:

\[
f(y,\rho) = \sum_{i=1}^Q W(y\, | \, x=i)^\frac{1}{1 + \rho} \tag{8} \label{eq:f}
\]

\[
g(y, \rho) = \frac{\partial}{\partial \rho} \left (f(y,\rho)^{1+\rho} \right ) \tag{9} \label{eq:g}
\]

We now find the derivatives of $E_0$ in terms of $f$ and $g$.  Note that we will use the  notation $f'(y,\rho) = \frac{\partial}{\partial \rho} f(y, \rho)$ since all derivatives are with respect to $\rho$.

We can rewrite $E_0(\rho)$ as 

\[
E_0(\rho) = (1+\rho)\log Q - \log \left( \int f(y, \rho)^{1 + \rho} \mathrm{d}y \right) \tag{10} \label{eq:E0_rewrite}
\] This yields the following for the derivatives of $E_0(\rho)$:

\[
E_0'(\rho) = \log Q - \frac{\int g(y, \rho) \text{d}y}{\int f(y, \rho)^{1 + \rho} \text{d}y} \tag{11} \label{eq:E0prime}
\]

\[
E_0''(\rho) = \frac{(\int g(y, \rho) \mathrm{d}y)^2}{(\int f(y, \rho)^{1 + \rho} \mathrm{d}y)^2} - \frac{\int g'(y, \rho) \mathrm{d}y}{\int f(y, \rho)^{1+\rho} \mathrm{d}y} \tag{12} \label{eq:E0primeprime}
\]

The relevant derivatives of $f$ and $g$ are shown in 
equations (\ref{eq:fprime})-(\ref{eq:gprime}).

\[
f'(y, \rho) = -\sum_{i=1}^Q \frac{W(y \, | \, i)^{\frac{1}{1 + \rho}} \log W(y \, | \, i)}{(1 + \rho)^2} \tag{13} \label{eq:fprime}
\]

\[
f''(y, \rho) = \sum_{i=i}^Q \frac{W(y \, | \, i)^{\frac{1}{1 + \rho}} \log W(y \, | \, i)}{(1 + \rho)^3} \left (\frac{\log W(y \, | \, i)}{1 + \rho} + 2 \right) \tag{14} \label{eq:fprimeprime}
\]

\[
g(y, \rho) = f(y, \rho)^{1 + \rho} \left (\log f(y, \rho) + \frac{(1 + \rho) f'(y, \rho)}{f(y, \rho)} \right ) \tag{15} \label{eq:gf}
\]

With these derivatives, we finally give the saddlepoint approximation for the RCU bound as given in \cite{Saddlepoint}.  Let $R = \frac{1}{n} \log M$ be the code rate.  We define $\hat{\rho}$ to be the unique solution to the equation $E_0'(\hat{\rho}) = R$.  We also define the \emph{channel dispersion} $V(\hat{\rho}) = -E_0''(\hat{\rho})$.  The saddlepoint approximation of the RCU bound is given by 

\[
\mathrm{rcu}(n, M) \simeq \tilde{\xi}_n(\hat{\rho}) + \psi_n(\hat{\rho}) e^{-n(E_0(\hat{\rho}) - \hat{\rho}R)} \tag{17} \label{eq:saddlepoint}
\] where the functions $\tilde{\xi}(.)$ and $\psi_n(.)$ are given by

\[
    \tilde{\xi}(\hat{\rho}) = \left\{
        \begin{array}{ll}
            1 \quad \quad \quad \quad \quad \quad \quad \quad \quad \hat{\rho} < 0 \\
            0 \quad \quad \quad \quad \quad \quad \quad \quad \quad 0 \le \hat{\rho} \le 1 \\
            e^{-n(E_0(1, P) - R)}\theta_n(1) \quad \, \hat{\rho} > 1
        \end{array}
    \right.
    \tag{18} \label{eq:xi}
\]

\[
\psi_n(\hat{\rho}) = \theta_n(\hat{\rho}) \left (\Psi(\hat{\rho} \sqrt{n V(\hat{\rho})}) + \Psi((1 \! - \! \hat{\rho})\sqrt{n V(\hat{\rho})} \right ) \tag{19} \label{eq:psi}
\]

The function $\psi_n(.)$ is given in terms of the functions $\Psi(.)$ and $\theta_n(.)$ which are defined as

\[
\Psi(z) = \frac{1}{2} \mathrm{erfc} \left (\frac{|z|}{\sqrt{2}} \right ) \exp \left (\frac{z^2}{2} \right ) \mathrm{sign}(z) \tag{20} \label{eq:Psi}
\]

\[
\theta_n(\hat{\rho}) \simeq \frac{1}{\sqrt{1 + \hat{\rho}}} \left (\frac{1 + \hat{\rho}}{\sqrt{2 \pi n \bar{\omega}''(\hat{\rho})}} \right )^{\hat{\rho}} \tag{21} \label{eq:theta}
\] where $\bar{\omega}''(\hat{\rho})$ is given as

\[
\bar{\omega}''(\hat{\rho}) = \int \mathcal{Q}_{\hat{\rho}}(y) \left [\frac{\partial^2}{\partial \tau^2} \alpha(y, \tau) \vline _{\tau = \hat{\tau}}   \right ] \mathrm{d}y \tag{22} \label{eq:omegaprimeprime}
\] and the derivative in $\bar{\omega}''(\hat{\rho})$ is evaluated at $\hat{\tau} = 1 / (1 + \hat{\rho})$.  The closed form expression for $\frac{\partial^2}{\partial \tau^2} \alpha(y, \tau)$ is given by equations (\ref{eq:alphaprimeprime})-(\ref{eq:betaprimeprime}).

\[
\frac{\partial^2}{\partial \tau^2} \alpha(y, \tau) = \frac{\beta(y, \tau) \frac{\partial^2}{\partial \tau^2} \beta(y, \tau) - \left (\frac{\partial}{\partial \tau} \beta(y, \tau) \right )^2}{  \beta(y, \tau)  ^2} \tag{23} \label{eq:alphaprimeprime}
\]

\[
\beta(y, \tau) = \frac{1}{M} \sum_{i = 1}^M W(y | i)^{\tau} \tag{24} \label{eq:beta}
\]

\[
\frac{\partial}{\partial \tau} \beta(y, \tau) = \frac{1}{M} \sum_{i = 1}^M W(y | i)^{\tau} \log W(y | i) \tag{25} \label{eq:betaprime}
\]

\[
\frac{\partial^2}{\partial \tau^2} \beta(y, \tau) = \frac{1}{M} \sum_{i = 1}^M W(y | i)^{\tau} \left (\log W(y | i) \right )^2 \tag{26} \label{eq:betaprimeprime}
\]

Finally, the distribution $\mathcal{Q}_{\hat{\rho}}(.)$ is defined as

\[
\mathcal{Q}_{\hat{\rho}}(y) = \frac{1}{\mu(\hat{\rho})} \left (\frac{1}{M} \sum_{i = 1}^M W(y | i)^{\frac{1}{1 + \hat{\rho}}} \right )^{1 + \hat{\rho}} \tag{27} \label{eq:Q}
\] where $\mu(\hat{\rho})$ is a normalization constant for the distribution $\mathcal{Q}_{\hat{\rho}}(y)$.

The noncoherent QFSK channel has the channel transition probability $W(y \, | \, x)$ given in Section \ref{sec:Channel}.  Noteworthy about this channel is that $y$ is a vector with length equivalent to the number of orthogonal signals, $Q$, and every element of $y$ is positive since they are generated from a Rayleigh or Rice distribution.  As such, the integrals with respect to $y$ in the saddlepoint approximation must be evaluated over $\mathbb{R}_{+}^Q$, the space of length-$Q$ real vectors with all positive elements.

While the calculation of the saddlepoint approximation of the RCU bound is less complex than the true calculation of the RCU bound, it is still very complex. The normal approximation \cite{Polyanskiy} relies on numerical integration
to find the channel capacity of the noncoherent QFSK channel \cite{Stark} and the channel dispersion $V$, and is thus much easier to calculate.
We now present the normal approximation for the QFSK/AWGN channel, following \cite{NormalApprox}.

Over the ensemble of length-$n$, rate-$R$ codes, according to the normal approximation the frame-error rate (FER) is approximately given by

\[
    FER = q\left(  \frac{n(C-R)\,+\,0.5\log_2(n)\,+\,O(1)}{\sqrt{nV}}    \right) \tag{28}
\]
where $q(x)=\int_{x}^{\infty}e^{-z^2/2}dz/\sqrt{2\pi}$ and $C$ and $R$ are both in units of bits per channel use. The channel capacity $C$ in this expression is given \cite{Stark} by the expectation of the channel information density, $i(X;Y)$, 

\[
C = \mathbb{E}\left[i(X;Y) \right] \tag{29}
\]
which can be rearranged \cite{Stark} into

\[
C = \log_2(Q)-\mathbb{E}_{y|x=1}\left[\log_2 \left(1 \,+\sum ^{Q}_{i=2}\Lambda_{i}(y)\,    \right)   \right]\text{ (bits/use)} \tag{30}
\]
where 


\[
\Lambda_{i}(y)=\frac{I_0\left(y_i\right)}{I_0\left(y_1\right)} \tag{31}
\]
and $y_1$ is Rician and each $y_i$ is Rayleigh.
In the FER expression, $V$ is the channel dispersion given by 

\[
V  = \mathbb{E}\left[\left(\log_2(Q)-\log_2 \left(1 \,+\sum ^{Q}_{i=2}\Lambda_{i}(y)\,
   \right)  - C\right)^2 \:\right] \tag{32}
\]

\section{Results}
\label{sec:Results}

We simulate CRC-ZTCC codes with the convolutional generator polynomials shown in Table \ref{tab:gen} and the CRC polynomials in Table \ref{tab:CRC}.  We used a message length of $K=64$ $4$-ary symbols, i.e. 128 bits, for all CRC-ZTCCs.  Our codes are rate-$K/(2*(K+m+\nu))$, where $m$ symbols are added for the CRC code and $\nu$ symbols are added for zero-state termination.  The decoder we used was an adaptive list Viterbi algorithm (LVA) decoder with a maximum list size of $2048$, as in \cite{King}.  The adaptive LVA employs parallel list decoding with an initial list size of 1 and doubles the list size until either a message candidate is found that passes the CRC check or the maximum list size of $2048$ is reached.

\begin{figure}
    \centering
    \includegraphics[width=20pc]{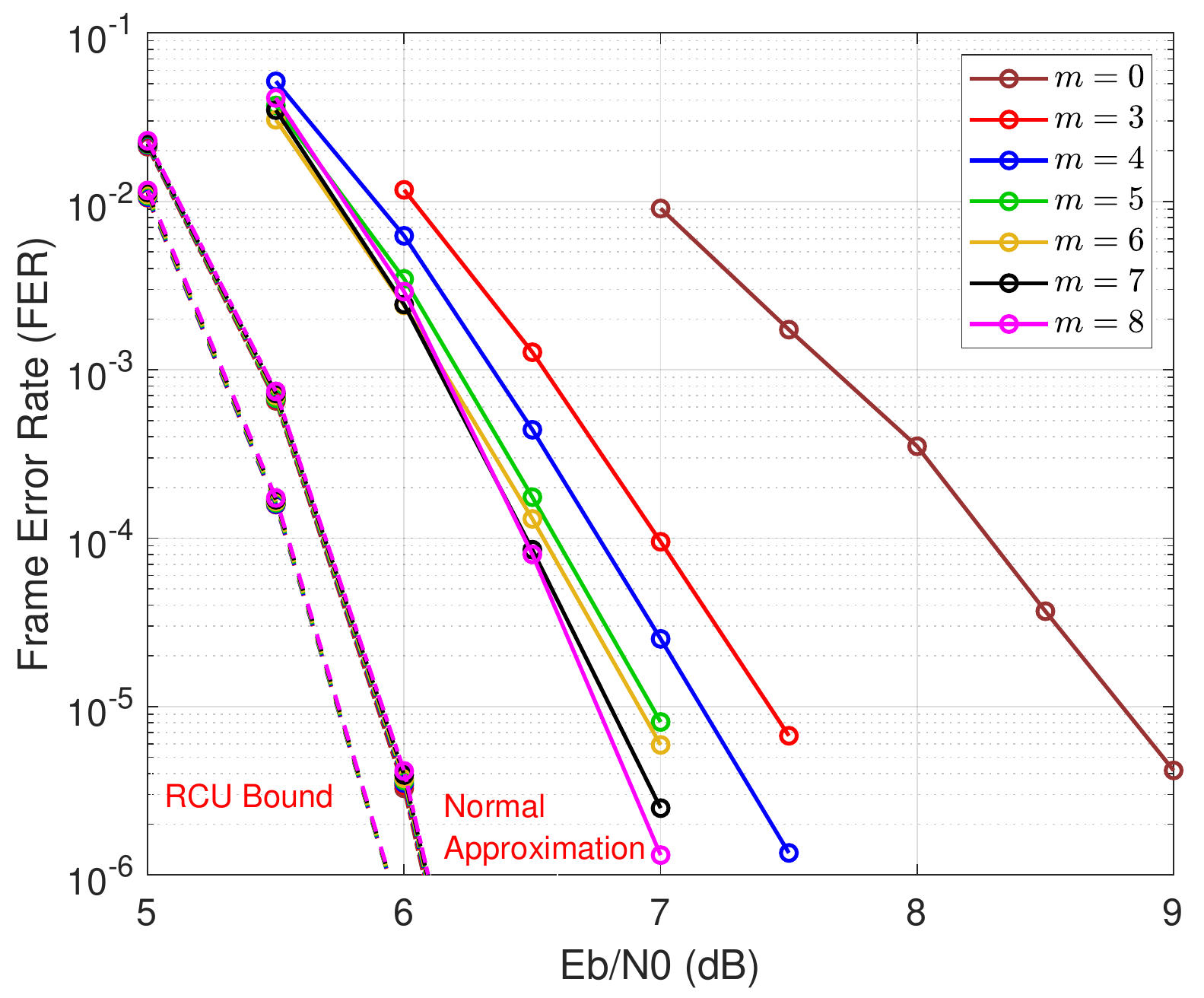}
    \caption{FER vs. $E_b/N_0$ for all $\nu=2$ CRC-ZTCC codes.  The solid lines are the data for codes, and the dashed lines are the corresponding RCU bounds.  Each message had a length of $K=64$ 4-ary symbols, and the decoder had a maximum list size of $2048$.  The best CRC-ZTCC code has a gap of about $0.9$ dB to RCU bound at an FER of $10^{-4}$.}
    \label{fig:Mem2}
\end{figure}

\begin{figure}
    \centering
    \includegraphics[width=20pc]{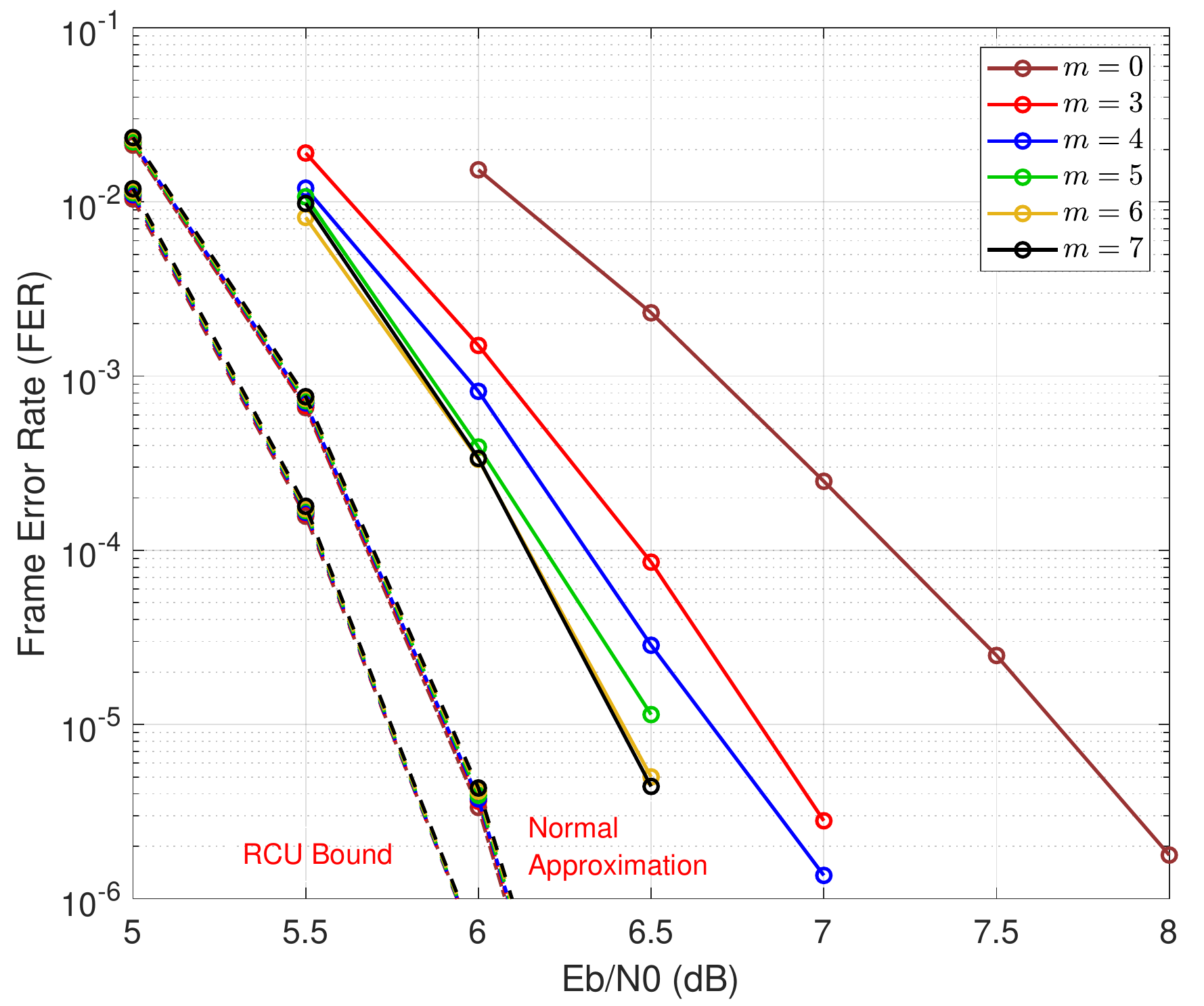}
    \caption{FER vs. $E_b/N_0$ for all $\nu=4$ CRC-ZTCC codes.  The solid lines are the data for codes, and the dashed lines are the corresponding RCU bounds.  Each message had a length of $K=64$ 4-ary symbols, and the decoder had a maximum list size of $2048$.  The best CRC-ZTCC code has a gap of about $0.59$ dB to RCU bound at an FER of $10^{-4}$.}
    \label{fig:Mem4}
\end{figure}

\begin{figure}
    \centering
    \includegraphics[width=20pc]{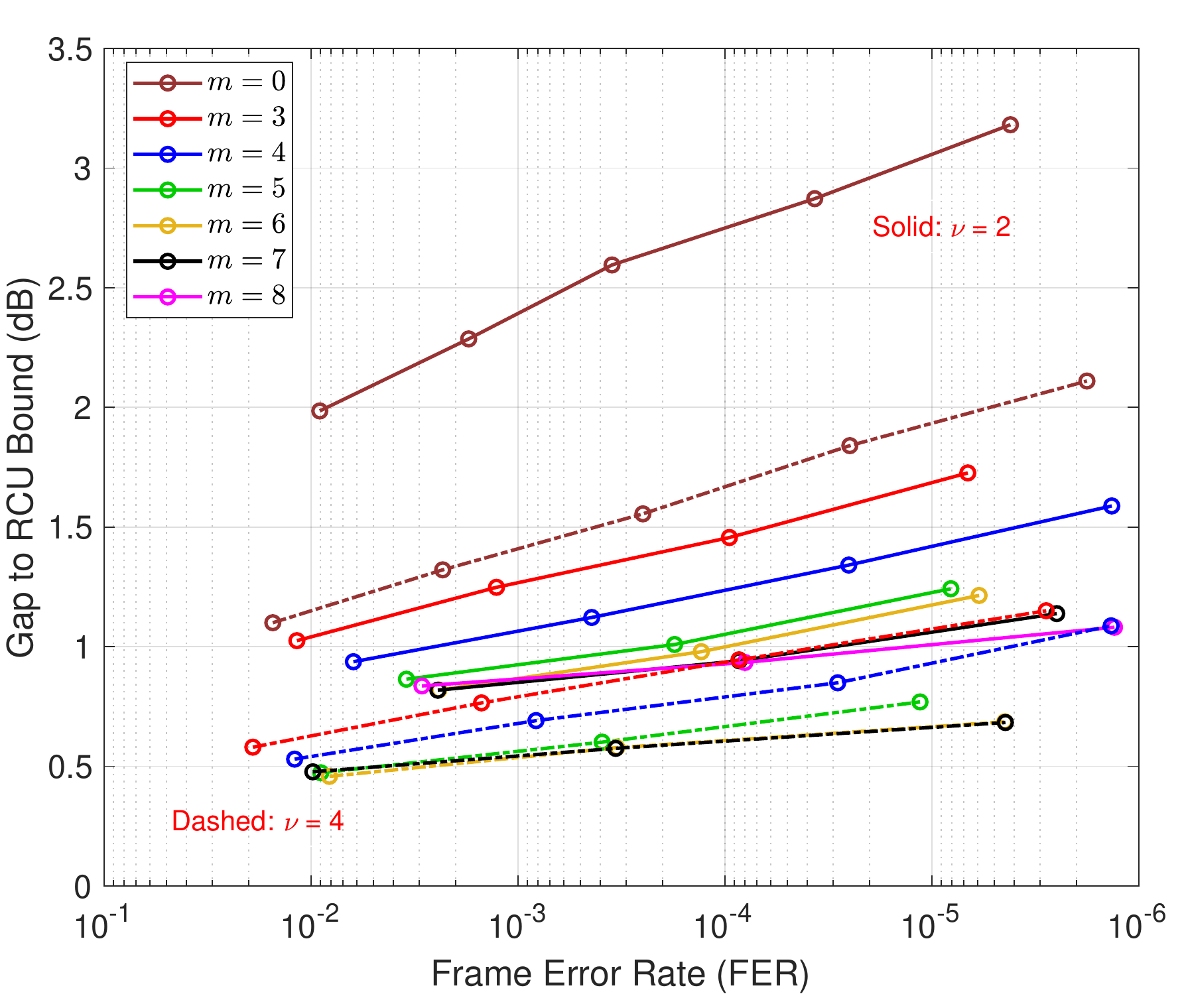}
    \caption{Gap to RCU bound vs. FER for every CRC-ZTCC code.  Higher values for $m$ have a smaller gap, and the $\nu=4$ codes have a smaller gap than the $\nu=2$ codes.  The gap also increases as FER decreases.}
    \label{fig:Gap}
\end{figure}

Fig. \ref{fig:Mem2} shows the FER vs. $E_b/N_0$ of the $\nu=2$ rate-1/2 CRC-ZTCC codes.  The values of $m$ vary from $m = 3$ to $m = 8$.  We also include the FER curve of the ZTCC without CRC concatenation ($m = 0$).  We see that increasing the length of the CRC improves the performance of the CRC-ZTCC code.  However, there is still a $1$ dB gap between the best FER performance and the RCU bound.

Fig. \ref{fig:Mem4} shows FER vs. $E_b/N_0$ for the $\nu=4$ rate-1/2 CRC-ZTCC codes. As in Fig. \ref{fig:Mem2}, the performance of the CRC-ZTCC improves as $m$ increases. Fig. \ref{fig:Mem4} shows $\nu=4$ CRC-ZTCCs approach the RCU bound more closely than the $\nu=2$ CRC-ZTCCs, reducing the gap down to 0.59 dB at $\textrm{FER} = 10^{-4}$.

Fig. \ref{fig:Gap} compares every code we have simulated to its respective RCU bound, to visualize the performance of all the codes, plotting the gap between each code's performance and the RCU bound as a function of the FER.  Larger values of $m$ achieve smaller gaps to the RCU bound. Generally, the $\nu=4$ CRC-ZTCCs have smaller gaps to the RCU bound than $\nu=2$ CRC-ZTCCs.   The gap to RCU bound increases as the FER decreases.

Our best 4-ary CRC-ZTCC has a gap to RCU bound of around $0.59$ dB at $\textrm{FER} = 10^{-4}$.  This matches closely with the results in \cite{Yang2022} for the analysis of binary CRC-ZTCCs.  This motivates the search for optimal CRCs for 4-ary tail biting convolutional codes (CRC-TBCCs), since binary CRC-TBCCs in \cite{Yang2022} approach the RCU bound closely.



Our simulations revealed that as $E_b/N_0$ decreases, the expected list size converges to around $4^m$, except when limited by our maximum list size of 2048.  This behavior matches the results in \cite{Yang2022}.   As $E_b/N_0$ increases, or as FER decreases, the expected list size converges to 1, also matching \cite{Yang2022}.  For example, at $E_b/N_0 = 6.5$ dB  the expected list size for all of the $\nu = 4$ CRC-ZTCCs is less than $1.06$.




\section{Conclusion}
\label{sec:Conclusion}

This paper presents CRC-ZTCC concatenated codes for $Q$-ary orthogonal signaling,  designing CRCs for specific ZTCCs to optimize the distance spectrum to achieve the best possible FER performance. To compare with our simulations, the paper also derives saddlepoint approximations of the RCU bound and presents the normal approximation for the noncoherent QFSK channel.

List decoding using a distance-spectrum optimal CRC significantly improves the minimum distance and the FER performance compared to the ZTCC decoded without the benefit of CRC-aided list decoding.  At FER $10^{-4}$, CRC-aided list decoding improves the $\nu=2$ ZTCC by between $1.2$ and $1.4$ dB  and the $\nu=4$ ZTCC by between $0.7$ and $0.8$ dB.  The performance improvement increases as the size of the CRC is increased.  At low FER (or equivalently high SNR) the average list size approaches 1 so that the average complexity burden of such list decoding is minimal.  Our best CRC-ZTCC design is within $0.59$ dB of the RCU bound at an FER of $10^{-4}$.

This paper focuses on ZTCC-CRC code designs, but TBCC-CRCs have been shown to approach very close to the RCU bound at short blocklengths.  Future work will design CRCs for $Q$-ary TBCCs.  Future work will also explore designs for higher values of $Q$.  This paper used sub-optimal decoding metrics for reduced complexity, and future investigations will assess the performance gain possible with optimal decoding metrics.

\bibliography{IEEEabrv, ref}

\end{document}